\documentstyle[12pt]{article}
\def\be{\begin{equation}}
\def\ee{\end{equation}}
\def\bea{\begin{eqnarray}}
\def\eea{\end{eqnarray}}

\begin{document}

\begin{center}
{\Large{\bf Actions for the Bosonic String with the Curved Worldsheet}}

\vskip .5cm
{\large Davoud Kamani}
\vskip .1cm
{\it Faculty of
 Physics, Amirkabir University of Technology (Tehran Polytechnic)\\
 P.O.Box: 15875-4413, Tehran, Iran}\\
{\it e-mail: kamani@cic.aut.ac.ir}\\
\end{center}

\begin{abstract}

At first we introduce an action for the string, which leads to a
worldsheet that always is curved. For this action we study the Poincar\'e
symmetry and the associated conserved currents. Then, a generalization
of the above action, which contains an arbitrary function of the
two-dimensional scalar curvature, will be introduced.
An extra scalar field enables us to modify these actions to Weyl invariant
models.

\end{abstract}
\vskip .5cm

{\it PACS}: 11.25.-w; 11.30.Cp

{\it Keywords}: Curved worldsheet; 2d scalar curvature; Poincar\'e symmetry.

\newpage
%%%%%%%%%%%%%%%%%%%%%%%%%%%%%%%%%%%%%%%%%%%%%%%%%%%%%%%%%%%%%%%%%%%%%%%%%%%
\section{Introduction}

The two-dimensional models have widely been used in the context of the
two-dimensional gravity ($e.g.$ see \cite{1,2,3,4} and references therein)
and string theory. From the 2d-gravity point of view, higher-dimensional
gravity models, by dimensional reduction reduce to the 2d-gravity
\cite{1,2,3}. From the string theory point of view, the (1+1)-dimensional actions
are fundamental tools of the theory. However, 2d-gravity and 2d- string theory
are closely related to each other. 

The known sigma models for string, in the presence of the dilaton field
$\Phi (X)$, contain the two-dimensional scalar curvature $R(h_{ab})$,
\bea
S_{\Phi} = \frac{1}{4\pi}\int d^2 \sigma \sqrt{h}R \Phi (X).
\eea
In two dimensions the combination
$\sqrt{h}R$ is total derivative. Thus, 
in the absence of the dilaton field, this action is a topological invariant
that gives no dynamics to the worldsheet metric $h_{ab}$. 

In fact, in the action (1), the dilaton is not the only choice. 
For example, replacing the dilaton field with the scalar curvature $R$,
leads to the $R^2$-gravity \cite{1,4,5}. In particular the Polyakov action
is replaced by a special combination of the worldsheet fields, which include an
overall factor $R^{-1}$.
Removing the dilaton and replacing
it with another quantities motivated us to study a class of 
two-dimensional actions. They are useful in the context of the
non-critical strings with curved worldsheet, and the
2-dimensional gravity.

Instead of the dilaton field, we introduce some combinations of $h_{ab}$,
$R$ and the induced metric on the worldsheet, $i.e.$ 
$\gamma_{ab}$, which give dynamics to $h_{ab}$.
These non-linear combinations can contain an arbitrary function $f(R)$ of the 
scalar curvature $R$. We observe that these dynamics lead to 
the constraint equation for $h_{ab}$, extracted from the Polyakov action. 

For the flat spacetime, these models have the Poincar\'e symmetry. 
In addition, they are reparametrization invariant. 
However, for any function $f(R)$,
they do not have the Weyl symmetry. Therefore, the string
worldsheet at most is conformally flat. By introducing an extra scalar
field in these actions, they also find the Weyl symmetry. Note that a
Weyl non-invariant string theory has noncritical dimension, 
$e.g.$ see \cite{6}.

This paper is organized as follows. In section 2, we introduce
a new action for the string in which the corresponding worldsheet
always is curved. In section 3, the Poincar\'e symmetry of this
string model will be studied. In section 4, the generalized form of the
above action will be introduced and it will be analyzed.
%%%%%%%%%%%%%%%%%%%%%%%%%%%%%%%%%%%%%%%%%%%%%%%%%%%%%%%%%%%%%%%%%%%%%%%%%%%%
\section{Curved worldsheet in the curved spacetime}

We consider the following action for the string, which propagates in the curved
spacetime  
\bea
S= -T \int d^2 \sigma \sqrt{h}R \bigg{(}R - 
\frac{1}{2\pi \alpha'} h^{ab}\gamma_{ab}\bigg{)},
\eea
where $h=-\det h_{ab}$, and $T$ is a dimensionless constant.
In addition, $R$ denotes the two-dimensional scalar curvature
which is made from $h_{ab}$. 
The string coordinates are $\{X^\mu (\sigma , \tau)\}$. 
The induced metric on the worldsheet, $i.e.$ $\gamma_{ab}$, is also given by
\bea
\gamma_{ab}= g_{\mu \nu}(X)\partial_a X^\mu (\sigma , \tau)
\partial_b X^\nu (\sigma , \tau) ,
\eea
where $g_{\mu \nu}(X)$ is the spacetime metric.

In two dimensions, the symmetries of the curvature tensor 
imply the identity 
\bea
R_{ab} - \frac{1}{2}h_{ab}R =0.
\eea
Therefore, the variation of the action (2) leads to the following
equation of motion for $h^{ab}$,
\bea
R_{ab} - \frac{1}{2\pi \alpha'} \gamma_{ab}=0.
\eea
This implies that the energy-momentum tensor, extracted from the action (2),
vanishes.

Contraction of this equation by $h^{ab}$ gives 
$R = \frac{1}{2\pi \alpha'} h^{ab}\gamma_{ab}$.
Introducing this equation and the equation (5) into (4) leads to 
\bea
T_{ab}^{({\rm Polyakov})} \equiv \gamma_{ab}
- \frac{1}{2}h_{ab}(h^{a'b'}\gamma_{a'b'})=0.
\eea
This is the constraint equation, extracted from the Polyakov action. Note that
the energy-momentum tensor, due to the action (2), 
is proportional to the left-hand-side of the equation (5). Thus, it
is different from (6). 

The equation of motion of the string coordinate $X^\mu (\sigma , \tau)$ also is
\bea
\partial_a (\sqrt{h}R h^{ab}\partial_b X^\mu)
+\sqrt{h}R h^{ab} \Gamma^\mu_{\nu \lambda}\partial_a X^\nu\partial_b X^\lambda =0.
\eea
Presence of the scalar curvature $R$ distinguishes this equation from
its analog, extracted from the Polyakov action.

Now consider those solutions of the equations of motion (5) and (7),
which admit constant scalar curvature $R$. For these solutions,
the equation (7) reduces to the equation of motion of the 
string coordinates, extracted from the Polyakov action
with the curved background. However,
for general solutions the scalar curvature $R$ depends on the worldsheet
coordinates $\sigma$ and $\tau$, and hence this coincidence does not occur.
%%%%%%%%%%%%%%%%%%%%%%%%%%%%%%%%%%%%%%%%%%%%%%%%%%%%%%%%%%%%%%%%%%%%%%%%%%%%
\subsection{The model in the conformal gauge}

Under reparametrization of $\sigma$ 
and $\tau$, the action (2) is invariant. That is,
in two dimensions the general coordinate transformations 
$\sigma \rightarrow \sigma'(\sigma , \tau)$ and 
$\tau \rightarrow \tau'(\sigma , \tau)$, depend on two free functions, namely
the new coordinates $\sigma'$ and $\tau'$. By means of such transformations
any two of the three independent components of $h_{ab}$ can be eliminated. 
A standard choice is a parametrization of the worldsheet such that
\bea
h_{ab}=e^{\phi (\sigma , \tau)} \eta_{ab} ,
\eea
where $\eta_{ab}=diag(-1 , 1)$, and $e^{\phi (\sigma , \tau)}$ is an unknown
conformal factor. The choice (8) is called the conformal gauge.
Since the action (2) does not have the Weyl symmetry (a local
rescaling of the worldsheet metric $h_{ab}$)
we cannot choose the gauge $h_{ab}=\eta_{ab}$.

The scalar curvature corresponding to the metric (8) is
\bea
R = -e^{-\phi}\partial^2 \phi ,
\eea
where $\partial^2 = \eta^{ab}\partial_a \partial_b$.
Thus, the action (2) reduces to
\bea
S' = -T \int d^2 \sigma e^{-\phi} \partial^2 \phi \bigg{(} 
\partial^2 \phi + \frac{1}{2\pi \alpha'}\eta^{ab}\gamma_{ab}\bigg{)}.
\eea 
According to the gauge (8),
this action describes a conformally flat worldsheet.
%%%%%%%%%%%%%%%%%%%%%%%%%%%%%%%%%%%%%%%%%%%%%%%%%%%%%%%%%%%%%%%%%%%%%%%%%%%%
\section{Poincar\'e symmetry of the model}

In this section we consider flat Minkowski space, 
$i.e.$ $g_{\mu\nu}(X)=\eta_{\mu\nu}$. Therefore, the equations of motion
are simplified to
\bea
R_{ab}-\frac{1}{2\pi \alpha'}\eta_{\mu\nu}
\partial_a X^\mu \partial_b X^\nu =0,
\eea
\bea
\partial_a (\sqrt{h}R h^{ab}\partial_b X^\mu)=0.
\eea

The Poincar\'e symmetry reflects the symmetry of the background
in which the string is propagating. It is described by the 
transformations
\bea
&~& \delta X^\mu = a^\mu_{\;\;\nu} X^\nu + b^\mu,
\nonumber\\
&~& \delta h^{ab} =0,
\eea 
where $a^\mu_{\;\;\nu}$ and $b^\mu$ are independent of the worldsheet coordinates
$\sigma$ and $\tau$, and $a_{\mu\nu}= \eta_{\mu \lambda}a^\lambda_{\;\;\nu}$
is antisymmetric. Thus, from the worldsheet point of view, these
transformations are global symmetries.
Under these transformations the action (2) is invariant. 
%%%%%%%%%%%%%%%%%%%%%%%%%%%%%%%%%%%%%%%%%%%%%%%%%%%%%%%%%%%%%%%%%%%%%%%%%%%%
\subsection{The conserved currents}

The Poincar\'e invariance of the action (2) is associated to the following 
Noether currents
\bea
&~& {\cal{J}}^{\mu\nu a} = \frac{T}{2\pi \alpha'}\sqrt{h}R
h^{ab} (X^\mu \partial_b X^\nu-X^\nu \partial_b X^\mu),
\nonumber\\
&~& {\cal{P}}^{\mu a} =\frac{T}{2\pi \alpha'}\sqrt{h}R
h^{ab}\partial_b X^\mu ,
\eea
where the current ${\cal{P}}^{\mu a}$ is corresponding to the translation
invariance and ${\cal{J}}^{\mu\nu a}$ is the current associated to
the Lorentz symmetry. 
According to the equation of motion (12) these are conserved currents
\bea
&~& \partial_a {\cal{J}}^{\mu\nu a} =0,
\nonumber\\
&~& \partial_a {\cal{P}}^{\mu a} =0.
\eea
%%%%%%%%%%%%%%%%%%%%%%%%%%%%%%%%%%%%%%%%%%%%%%%%%%%%%%%%%%%%%%%%%%%%%%%%%%%%
\subsection{The covariantly conserved currents}

It is possible to construct two other currents from (14), in which they be
covariantly conserved.
For this, there is the useful formula 
\bea
\nabla_a K^a = \frac{1}{\sqrt{h}}\partial_a (\sqrt{h}K^a),
\eea
where $K^a$ is a worldsheet vector.
Therefore, we define the currents $J^{\mu \nu a}$ and $P^{\mu a}$ as
in the following
\bea
&~& J^{\mu\nu a}=\frac{1}{\sqrt{h}}{\cal{J}}^{\mu\nu a},
\nonumber\\
&~& P^{\mu a}=\frac{1}{\sqrt{h}}{\cal{P}}^{\mu a}.
\eea
According to the equations (15) and (16),
these are covariantly conserved currents, $i.e.$,
\bea
\nabla_a J^{\mu\nu a}= \nabla_a P^{\mu a}=0.
\eea
The currents (17) can also be written as
\bea
&~& J^{\mu\nu}_a = \frac{T}{2\pi \alpha'}R
(X^\mu \partial_a X^\nu-X^\nu \partial_a X^\mu),
\nonumber\\
&~& P^{\mu}_a =\frac{T}{2\pi \alpha'}R\partial_a X^\mu .
\eea
Since there is $\nabla_a h_{bc}=0$, the conservation laws (18) 
also imply the covariantly conservation of the currents (19).
%%%%%%%%%%%%%%%%%%%%%%%%%%%%%%%%%%%%%%%%%%%%%%%%%%%%%%%%%%%%%%%%%%%%%%%%%%%%
\section{Generalization of the model}

The generalized form of the action (2) is
\bea
I= -T \int d^2 \sigma \sqrt{h}R \bigg{(}f(R) - 
\frac{1}{2\pi \alpha'} h^{ab}\gamma_{ab}\bigg{)},
\eea
where $f(R)$ is an arbitrary differentiable function of the scalar
curvature $R$. The set $\{X^\mu(\sigma , \tau)\}$ describes a string
worldsheet in the spacetime. These string coordinates appeared in
the induced metric
$\gamma_{ab}$ through the equation (3). Thus, (20) is a model for the
string action.

The equation of motion of $X^\mu$ is as previous, $i.e.$ (7).
Vanishing the variation of this action with respect to
the worldsheet metric $h^{ab}$, gives the equation of motion of $h^{ab}$,
\bea
R_{ab}\frac{df(R)}{dR} - \frac{1}{2\pi \alpha'} \gamma_{ab}=0.
\eea
The trace of this equation is
\bea
R\frac{df(R)}{dR} - \frac{1}{2\pi \alpha'}h^{ab} \gamma_{ab}=0.
\eea
Combining the equations (4), (21) and (22) again leads to the equation (6).

As an example, consider the function $f(R) = \alpha \ln R + \beta$. 
Thus, the field
equation (21) implies that the intrinsic metric $h_{ab}$ becomes proportional
to the induced metric $\gamma_{ab}$, that is 
$h_{ab} = \frac{1}{\pi \alpha \alpha'}\gamma_{ab}$.

Since the Poincar\'e transformations contain $\delta h^{ab}=0$,
the generalized 
action (20) for the flat background metric $g_{\mu\nu}=\eta_{\mu\nu}$, 
also has the Poincar\'e invariance. This leads to the previous 
conserved currents, $i.e.$ (14) and (19).
%%%%%%%%%%%%%%%%%%%%%%%%%%%%%%%%%%%%%%%%%%%%%%%%%%%%%%%%%%%%%%%%%%%%%%%%%%%%%%%
\subsection{Weyl invariance in the presence of a new scalar field}

The action (20) under the reparametrization transformations is symmetric.
The Weyl transformation is also defined by
\bea
h_{ab} \longrightarrow h'_{ab}=e^{\rho (\sigma , \tau)}h_{ab}.
\eea
Thus, the scalar curvature transforms as
\bea
R \longrightarrow R' = e^{-\rho}(R -\nabla^2 \rho),
\eea
where $\nabla^2 \rho = \frac{1}{\sqrt{h}}\partial_a (\sqrt{h}h^{ab}\partial_b\rho)$.
The equations (23) and (24) imply that the action (20), for any function $f(R)$, is
Weyl non-invariant. 

Introducing (23) and (24) into the action (20) gives a new action 
which contains the field $\rho (\sigma , \tau)$,
\bea
I' = -T \int d^2 \sigma \sqrt{h}(R- \nabla^2 \rho)
\bigg{(}f[e^{-\rho}(R- \nabla^2 \rho)] 
-\frac{1}{2\pi \alpha'}e^{-\rho}h^{ab} \gamma_{ab}
\bigg{)}.
\eea
We can ignore the origin of this action. 
In other words, it is another model for string.
However, under the Weyl transformations
\bea
&~& h_{ab} \longrightarrow e^{u (\sigma , \tau)}h_{ab},
\nonumber\\
&~& \rho \longrightarrow \rho - u ,
\eea
the action $I'$, for any function $f$, is symmetric. Note that according
to the definition of $\nabla^2$ there is the
transformation $\nabla^2 \rightarrow e^{-u}\nabla^2$. 
%%%%%%%%%%%%%%%%%%%%%%%%%%%%%%%%%%%%%%%%%%%%%%%%%%%%%%%%%%%%%%%%%%%%%%%%%%%
\section{Conclusions}

We considered some string actions which give dynamics to the worldsheet
metric $h_{ab}$. Due to the absence of the Weyl invariance, 
these models admit at most conformally flat (but not flat) worldsheet.
We observed that the constraint
equation on the metric, extracted from the Polyakov action, is a 
special result of the field equations of our string models. Obtaining this
constraint equation admits us to introduce an arbitrary function
of the scalar curvature to the action. 
For the case $f(R) = \alpha \ln R + \beta$, the metric $h_{ab}$ becomes
proportional to the induced metric of the worldsheet.

By introducing 
a new degree of freedom we obtained a string action, in which for any
function $f$ is Weyl invariant. 

Our string models with arbitrary $f(R)$, in the flat background 
have the Poincar\'e symmetry.
The associated conserved currents are proportional to the scalar
curvature $R$. We also constructed the covariantly conserved currents
from the Poincar\'e currents.
%%%%%%%%%%%%%%%%%%%%%%%%%%%%%%%%%%%%%%%%%%%%%%%%%%%%%%%%%%%%%%%%%%%%%%%%%%%%

\end{document}